\documentstyle[12pt,draft,epsf]{article}
\newcommand{\resection}[1]{\setcounter{equation}{0}\section{#1}}

\def\l{\lambda}

\def\be{\begin{equation}}
\def\ee{\end{equation}}
\def\bea{\begin{eqnarray}}
\def\eea{\end{eqnarray}}
\def\beano{\begin{eqnarray*}}
\def\eeano{\end{eqnarray*}}
\def\bd{\begin{displaystyle}}
\def\ed{\end{displaystyle}}
\def\ba{\begin{array}}
\def\ea{\end{array}}

\begin{document}
\oddsidemargin 5mm
\setcounter{page}{0}
\newpage     
\setcounter{page}{0}
\begin{titlepage}
\begin{flushright}
IFT-P.067/2002\\
CLNS 02/1798\\
\end{flushright}
\vspace{0.5cm}
\begin{center}
{\large {\bf Noncommutative Integrable Field Theories in 2d}}\\
\vspace{1.5cm}
{\bf I. Cabrera-Carnero} \footnote{\tt{email:cabrera@ift.unesp.br}}\\
{\em IFT, Unesp}\\
{\em Rua Pamplona, 145}\\
{\em S\~ao Paulo, SP 01405-900, Brazil}\\
\vspace{0.8cm}
{\bf M. Moriconi} \footnote{\tt{email:marco@if.ufrj.br}}\\
{\em Newman Laboratory of Nuclear Studies, Cornell University}\\
{\em Ithaca, New York 14853, USA}\\
{\rm and}\\
{\em Instituto de F\'\i sica}\\
{\em Universidade Federal do Rio de Janeiro}\\
{\em Rio de Janeiro, RJ 21945-970, Brazil}\\
\vspace{0.8cm}
\end{center}
\renewcommand{\thefootnote}{\arabic{footnote}}
\vspace{6mm}

\begin{abstract}
\noindent
We study the noncommutative generalization of (euclidean) integrable
models in two-dimensions, specifically the sine- and sinh-Gordon and
the $U(N)$ principal chiral models. By looking at tree-level
amplitudes for the sinh-Gordon model we show that its na\"\i ve
noncommutative generalization is {\em not} integrable. On the other hand, the
addition of extra constraints, obtained through the generalization of
the zero-curvature method, renders the model integrable. We construct
explicit non-local non-trivial conserved charges for the $U(N)$
principal chiral model using the Brezin-Itzykson-Zinn-Justin-Zuber method.
\vspace{3cm}

%PACS number(s): 11.55, 05.50+q  
\end{abstract}
\vspace{5mm}
\end{titlepage}

\newpage
%\baselineskip=24pt
%\pagebreak
\setcounter{footnote}{0}
\renewcommand{\thefootnote}{\arabic{footnote}}

\resection{Introduction} 

Noncommutative field theories (ncft's)have attracted a great deal of
attention recently, due to their relation to string theory, where they
arise as a limit of type IIB theories with a B-field turned on \cite{SW}.
Besides this important connection, ncft's are interesting on their own
setting, with a very rich and unexpected structure, such as the UV/IR
mixing for example \cite{MvRS}, and applications to the quantum
Hall effect \cite{Suss}.

It has been shown in general that the introduction of space-time
noncommutativity leads to non-unitary theories \cite{GM}, but it is
conceivable that some specific models could evade some of these arguments
\cite{CLZ,BDFP}. Since in a noncommutative theory in two-dimensions we
necessarily have space-time noncommutativity, we have to be careful in
defining the theory properly. One way to avoid these complications is to
consider two-dimensional {\em euclidean} models.

We argue that after introducing noncommutativity, obtained by considering
the replacement of the product of the fields in the action by their
$\star$-products, some of these models are still integrable classically,
whereas others are not. We show that models obtained in this way that are
not integrable, can be redefined by a suitable generalization of the
zero-curvature method \cite{FT} and then shown to be integrable.

This paper is organized as follows. In the next section we briefly
review perturbative non-commutative field theory. In section 3 we
discuss some of the generalities of integrable field theories,
introduce the models we are going to study, show the non-integrability
of the noncommutative sG and shG models, discuss the noncommutative
generalization of the zero-curvature formalism, and show how the
integrability of the sG and shG models may be restored and present
soliton (localized) solutions. We also discuss the $U(N)$ pcm and show
that its noncommutative generalization is integrable. In this case we
construct non-local charges following the method of \cite{BIZZ}. In section
4 we present our conclusions and comment on future directions to
pursue. Some of the technical aspects are presented in the appendices.

\section{Non-Commutative Field Theories}

Let us consider scalar field theories for simplicity.  We construct a
ncft \cite{reviews} from a given quantum field theory (qft) by replacing
the product of fields by the $\star$-product
\be
\phi_1(x)\phi_2(x) \to 
\phi_1(x)\star\phi_2(x)=e^{\frac{i}{2}\theta^{\mu\nu}\partial_\mu^{x_1}
\partial_\nu^{x_2}}\phi_1(x_1)\phi_2(x_2)|_{x_1=x_2=x}
\ee
This deformation of the usual product implies in a change in the
Feynman rules.  We refer the reader to Filk's paper \cite{Filk} for a
more complete discussion of Feynman rules in ncft (see also
\cite{MvRS}, \cite{reviews}). 
Here we review the essential aspects to our discussion.

A simplifying aspect in the analysis of ncft's is that the propagator of a 
ncft is the same as the one of its commuting
version. This is due to the fact that, for a manifold without boundaries,
\be
\int dx f(x) \star g(x)=\int dx f(x) g(x)
\ee
Therefore, the quadratic part of the action is the same for the
noncommutative version of the model, providing the same propagator.

In the following we will refer to functions of operators in the
noncommutative deformation by a $\star$ sub-index, for example
$\phi^{n}_\star=\phi\star\phi\star \ldots \star \phi$.

If on one hand propagators are the same as in the commutative versions, 
vertices will pick up phases. For example, if we
consider a $\phi^n_\star$ term in a two-dimensional
scalar field theory, we obtain in momentum space
\be
\int dx \phi(x) \star \ldots \star \phi(x) =
\int \prod_{i=1}^{n}dp_i e^{-\frac{i}{2}\sum_{k<m}(p_k)_\mu \theta^{\mu \nu}(p_m)_\nu}
\tilde{\phi}(p_1)\ldots\tilde{\phi}(p_n) \delta(p_1+\ldots+p_n)
\ee

Notice that already at tree-level, there will be differences in the
scattering amplitudes of a commutative theory and its noncommutative
counterpart, since the vertices are modified.

Let us see what are the changes in the cases that will be of interest to us,
namely the 4- and 6-point vertices in a scalar theory. The
4-point vertex changes according to
\be
\int dx \phi^4_{\star}=\int dx (\exp(-i\sum_{i<j}k_i \wedge k_j))\phi(k_1)
\phi(k_2)\phi(k_3)\phi(k_4) \delta(k_1+k_2+k_3+k_4) \ , \ \label{v4}
\ee
where we introduced the notation $k_i \wedge k_j = \frac{1}{2}((k_i)_\mu \theta^{\mu \nu} (k_j)_\nu)$.
We should remark that in two dimensions $k_i \wedge k_j = \frac{\theta}{2}((k_i)_1(k_j)_2-(k_i)_2(k_j)_1)$, since
$\theta^{\mu \nu} = \theta \epsilon^{\mu \nu}$.
In general the Moyal deformation of vertices does not preserve the
permutation symmetry, but in the case of a single scalar boson, we can
actually symmetrize the integrand, and replace the phases in \ref{v4}
by
\bea
&&G_4(k_1,k_2,k_3,k_4)=\frac{1}{4!}
\sum_{perm.} \exp(-i\sum_{i<j} k_i \wedge k_j)=
\frac{1}{3}(\cos(k_1\wedge k_2) \cos(k_3 \wedge k_4) +\\
&&\phantom{G_4(k_1,k_2,k_3,k_4)=}
\cos(k_1\wedge k_3) \cos(k_2 \wedge k_4)+
\cos(k_1\wedge k_4)\cos(k_2 \wedge k_3)) \ , \ 
\eea 
The analysis of the 6-point vertex is very similar, and gives
\be
G_6(k_1,k_2,k_3,k_4,k_5,k_6)=
\frac{1}{6!}\sum_{perm.}\exp(-i\sum_{i<j}k_i \wedge k_j) \label{v6}
\ee
We will leave the 6-point vertex in this form, since there is no simpler
way to write it, as in the case of the 4-point vertex.
All one has to do in order to compute amplitudes in a 
noncommutative scalar field theory is to write down exactly the same
Feynman graphs as in the commutative theory and replace the vertices
by expressions like \ref{v4} and \ref{v6} (and their analogous for higher
order vertices).

\section{Noncommutative Integrable Field Theories}

The existence of non-trivial higher-spin conserved charges has
dramatic consequences in the dynamics of two-dimensional qft's: there
is no particle production in a scattering process, the set of in and
out momenta is the same, multi-particle amplitudes are factorized into
products of two-body processes, and the two-body $S$-matrix satisfy
the Yang-Baxter equation, besides the usual analiticity and
crossing-symmetry properties (\cite{ZZ}, \cite{D}). 

We will consider the noncommutative extensions of the sine- and
sinh-Gordon (sG and shG) models and of the $U(N)$ principal chiral
model (pcm). The noncommutative sG model was studied in \cite{NOS} in
the context of $S$-duality, and its relation to the noncommutative
Thirring model through noncommutative bosonization, and the pcm was
studied in \cite{P} (see also \cite{LP}). In the sG and shG models the na\"\i ve Moyal
deformation leads to non-integrable field theories, whereas in the
case of the $U(N)$ pcm, integrability is preserved. On the other hand
it is possible to introduce new constraints in the sG and shG models
in such a way to restore integrability.

\subsection{Noncommutative sine- and sinh-Gordon}

The lagrangian of the shG model is 
\be 
{\cal L}_{shG}=\frac{1}{2}(\partial \phi)^2
+\frac{m^2}{\beta^2}(1-\cosh(\beta \phi)) \ . \ \label{shG}
\ee
As it is well known, the shG model is integrable, and is related to
the sG model, by replacing $\beta \to i\beta$.  
The equation of motion of the shG model can be easily derived from
\ref{shG} to be
\be
\frac{\partial^2 \phi}{\partial t^2}-\frac{\partial^2 \phi}{\partial x^2}+
\frac{m^2}{\beta}\sinh(\beta \phi)=0
\ee
One may be tempted at guessing that the noncommutative version of the
shG model, obtained by replacing the products of local fields in the
action by $\star$-products, would lead to an integrable model. This
turns out not to be the case.  Notice that the classical
non-integrability of the shG model implies the same for the sG model.

Consider the Moyal deformation of the shG lagrangian
\be
{\cal L}_{shG}^{\star}= \frac{1}{2}(\partial \phi)^2
+\frac{m^2}{\beta^2}(1-\cosh(\beta\phi)_{\star}) \ . \ \label{shGstar}
\ee
where
\be
\cosh(\beta \phi)_{\star}=\sum_{n=0}^{\infty}\frac{1}{(2n)!}
\left(\frac{\beta \phi}{2}\right)^{2n}_{\star}
\ee
The corresponding equation of motion is
\be
\frac{\partial^2 \phi}{\partial t^2}-\frac{\partial^2 \phi}{\partial x^2}+
\frac{m^2}{\beta}\sinh(\beta \phi)_{\star}=0 \label{eomstar}
\ee

The fact that the amplitude for particle production processes vanishes
exactly in an integrable model means that they vanish to each
order in a loop expansion, that is, in powers of $\hbar$. In
particular it should vanish at tree-level, which corresponds to the
classical limit of the theory, and is the hallmark of classical
integrability. Therefore, if any tree-level amplitude for a particle
production process is non-zero, we may be sure that this model is not
classically integrable. This strategy of showing non-integrability for
a given model was used in \cite{Sh}.

In the following we compute the tree-level amplitude for $2 \to 4$
particles and show that it vanishes in the shG model (see \cite{D}),
but that it does {\em not} vanish in its na\"\i ve noncommutative
deformation \footnote{In this subsection we compute the amplitudes in 
Minkowski space, since there is no problem with unitarity at tree-level.}. 
For this specific computation we need only to consider the
truncated lagrangian
\be
{\widetilde {\cal L}}_{shG}=
\frac{1}{2}(\partial \phi)^2 -\frac{m^2}{2}\phi^2-\frac{m^2\beta^2}{4!}\phi^4
-\frac{m^2\beta^4}{6!}\phi^6
\ee
Let us denote the in-momenta $p_1$ and $p_2$, and the out-momenta
$p_3,p_4, p_5$, and $p_6$. The amplitude for $p_1+p_2 \to
p_3+p_4+p_5+p_6$ will be denoted by ${\cal M}_{2 \to
4}=(2\pi)^4 \delta(p_1+p_2-p_3-p_4-p_5-p_6)
T(p_1,p_2;p_3,p_4,p_5,p_6)$.  Using the rapidity variable we can
write the in- and out-momenta as
$p_i=m(\cosh(\theta_i),\sinh(\theta_i))$.  In light-cone coordinates
it becomes $p_i^{\pm}=p_i^0\pm p_i^1= m \exp(\pm \theta_i)$. We will
denote the numbers $\exp(\theta_i)=a_i$. In the following we will consider
$T$ alone. 

The amplitude $T(p_1,p_2;p_3,p_4,p_5,p_6)$ gets contributions from three
types of diagrams, as shown in figure 1, where we still have to sum
over possible permutations of the in- and out-lines.
\vskip 0.5cm
\centerline{\epsffile{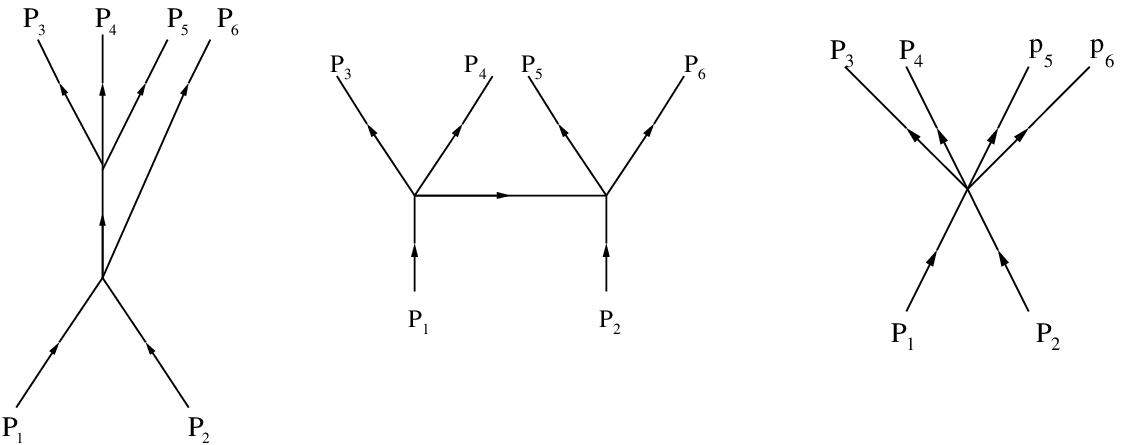}}
\vskip 0.5cm
\centerline{Fig. 1 The three types of diagrams contributing to $T$}
\vskip 0.5cm
We will call the amplitude for the first type of diagrams
$A(p_1,p_2;p_3,p_4,p_5,p_6)$ and for the second
$B(p_1,p_2;p_3,p_4,p_5,p_6)$. The third type is simply the $\phi^6$
vertex. It is easy to see that
\bea
A(p_1,p_2;p_3,p_4,p_5,p_6)=\frac{1}{(p_1+p_2-p_6)^2-m^2}=
-\frac{1}{m^2}\frac{a_1a_2a_6}{(a_1+a_2)(a_1-a_6)(a_2-a_6)} 
\nonumber \\
B(p_1,p_2;p_3,p_4,p_5,p_6)=\frac{1}{(p_1-p_3-p_4)^2-m^2}=
-\frac{1}{m^2}\frac{a_1a_3a_4}{(a_1-a_3)(a_1-a_4)(a_3+a_4)}
\eea
The final scattering amplitude has factors that depend on the external
legs, which are the same for all diagrams, and therefore unimportant
in our computation.

The scattering amplitude for the $2 \to 4$ process is, therefore, 
proportional to
\bea
&&T(p_1,p_2;p_3,p_4,p_5,p_6)\,\,=
A(p_1,p_2;p_3,p_4,p_5,p_6)+A(p_1,p_2;p_4,p_5,p_6,p_3)+\nonumber \\
&&A(p_1,p_2;p_5,p_6,p_3,p_4)+A(p_1,p_2;p_6,p_3,p_4,p_5)+B(p_1,p_2;p_3,p_4,p_5,p_6)+\nonumber \\
&&B(p_1,p_2;p_3,p_5,p_4,p_6)+B(p_1,p_2;p_3,p_6,p_5,p_4)+B(p_1,p_2;p_5,p_6,p_3,p_4)+\nonumber \\
&&B(p_1,p_2;p_4,p_6,p_3,p_5)+B(p_1,p_2;p_5,p_4,p_3,p_6)+1 \label{t1}
\eea
where the $1$ is the contribution from the 6-point vertex. By using 
energy-momentum conservation, which corresponds to
$a_1+a_2=a_3+a_4+a_5+a_6$ and
$1/a_1+1/a_2=1/a_3+1/a_4+1/a_5+1/a_6$, it can be shown that the above
expression vanishes! In particular this means that the contribution coming
from the 4-point vertices (amplitudes $A$ and $B$) add up to a 
constant (-1), and the constant contribution from the 6-point vertex (+1) 
precisely cancels it.

Let us consider the noncommutative amplitude now. 
Using formula \ref{v4} the noncommutative amplitudes $\tilde A$ 
and $\tilde B$ become
\bea
&&{\widetilde A}(p_1,p_2;p_3,p_4,p_5,p_6)=
A(p_1,p_2;p_3,p_4,p_5,p_6 )G_4(p_1,p_2,p_1+p_2-p_6,p_6) \nonumber \\
&&\phantom{{\widetilde A}(p_1,p_2;p_3,p_4,p_5,p_6)=}
G_4(p_1+p_2-p_6,p_3,p_4,p_5)
\nonumber \\
&&{\widetilde B}(p_1,p_2;p_3,p_4,p_5,p_6)=
B(p_1,p_2;p_3,p_4,p_5,p_6) G_4(p_1,p_3,p_4,p_1-p_3-p_4) \nonumber \\
&&\phantom{{\widetilde B}(p_1,p_2;p_3,p_4,p_5,p_6)=}
G_4(p_1-p_3-p_4,p_2,p_5,p_6)
\eea
the amplitude for the $2 \to 4$ process, is now 
\bea
&&{\widetilde T}(p_1,p_2;p_3,p_4,p_5,p_6)=
{\widetilde A}(p_1,p_2;p_3,p_4,p_5,p_6)+{\widetilde A}(p_1,p_2;p_4,p_5,p_6,p_3)+\nonumber \\
&&{\widetilde A}(p_1,p_2;p_5,p_6,p_3,p_4)+{\widetilde A}(p_1,p_2;p_6,p_3,p_4,p_5)+
{\widetilde B}(p_1,p_2;p_3,p_4,p_5,p_6)+ \nonumber \\
&&{\widetilde B}(p_1,p_2;p_3,p_5,p_4,p_6)+
{\widetilde B}(p_1,p_2;p_3,p_6,p_5,p_4)+{\widetilde B}(p_1,p_2;p_5,p_6,p_3,p_4)+\nonumber \\
&&{\widetilde B}(p_1,p_2;p_4,p_6,p_3,p_5)+{\widetilde B}(p_1,p_2;p_5,p_4,p_3,p_6)+
G_6(p_1,p_2;p_3,p_4,p_5,p_6) \label{t2}
\eea
where $G_6$ is given by \ref{v6}.
Once again, we have to take into account the energy-momentum
conservation constraint in evaluating this expression. As expected,
the zeroth order in $\theta$ is the amplitude \ref{t1}, and therefore
it vanishes. On the other hand, the expression \ref{t2} does not
vanish to the next order in $\theta$ (which is actually $\theta^2$),
and so this model is {\em not} integrable: the Moyal deformation 
of the shG model is not integrable. This does not mean
that there is no noncommutative version of the shG and sG models, but only
that our first attempt does not work. We will see now how to define
the noncommutative shG and sG models in such a way to obtain integrable
theories that reduce to the appropriate limits as $\theta \to 0$.

\subsection{Zero-Curvature Condition}

The definition of the noncommutative shG and sG  models as the Moyal
deformation of the action of their actions does not give an
integrable field theory. On the other hand we can define the noncommutative 
sG and shG models through the noncommutative generalization of the zero-curvature 
condition, which will provide, by construction, a theory with an infinite number
of conserved charges, and gives the usual sG and shG models
in the limit $\theta \to 0$. We start by reviewing the zero-curvature
method.

The equations of motion of an integrable field theory
in two dimensions can be written in the form
\be
\frac{\partial U}{\partial t}-\frac{\partial V}{\partial x}+[U,V]=0 \ , \
\label{zc}
\ee 
where $U$ and $V$ are two given potentials, which depend on space and
time, and a spectral parameter $\lambda$, and $[U,V]=UV-VU$. This is
the so-called zero-curvature condition, which encodes the integrable structure
of the theory. It corresponds to the
compatibility of the following pair of differential equations
\be
\frac{\partial F}{\partial x}=U F
\qquad {\rm and} \qquad
\frac{\partial F}{\partial t}=V F
\ee
where $F$ is an auxiliary vector.
We introduce now the $\star$-zero-curvature condition. Similarly to the usual 
zero-curvature condition, the $\star$-zero curvature
condition arises from the compatibility of the following pair of
differential equations
\be
\frac{\partial F}{\partial x}=U \star F
\qquad {\rm and} \qquad
\frac{\partial F}{\partial t}=V \star F \label{starzc}
\ee
We will show now, that the \ref{starzc} implies the existence
of an infinite number of conserved charges.

We will consider our theory to be defined on the interval $[-L,L]$,
and that the operators $U$ and $V$ satisfy periodic boundary conditions.
The equation satisfied by the monodromy operator $T_\l(x)$ is
\be
\frac{\partial T_\l}{\partial x} = U \star \label{monodromy}
T_\l 
\ee
with the boundary condition $T_\l(-L)= 1$. The solution of \ref{monodromy}
is easily seen to be
\be
T_\l(x)=P_\star\exp\left(-\int_{-L}^x dz \,\, U(z;\l)\right)
\ee
where $P_\star$ is the $\star$-path-ordered operator.
By taking the time derivative of \ref{monodromy} and using the
$\star$-zero-curvature condition, we obtain
\bea
&&\frac{\partial^2 T_\l }{\partial_x\partial_t} =
\frac{\partial U}{\partial t} \star T_\l+
U\star \frac{\partial T_\l}{\partial t} \nonumber \\
&&\phantom{\frac{\partial^2 T_\l}{\partial_x\partial_t}}=
\frac{\partial V}{\partial x} \star T_\l -[U,V]_\star \star T_\l + 
U \star \frac{\partial T_\l}{\partial t} \nonumber \\
&&\phantom{\frac{\partial^2 T_\l}{\partial_x\partial_t}}=
\frac{\partial V}{\partial x} \star T_\l-U \star V \star T_\l+
V\star \frac{\partial T_\l}{\partial x}+U\star \frac{\partial T_\l}{\partial t}
\nonumber \\
&&\phantom{\frac{\partial^2 T_\l}{\partial_x\partial_t}}=
\frac{\partial }{\partial x} (V \star T_\l)+U \star 
\left(\frac{\partial T_\l}{\partial t} -V \star T_\l \right)
\eea
which implies
\be
\frac{\partial}{\partial x}
\left(\frac{\partial T_\l}{\partial t}-V\star T_\l \right)=
U \star \left(\frac{\partial T_\l}{\partial t}-V\star T_\l \right)
\ee
This means that
\be
\frac{\partial T_\l(x)}{\partial t}= V(x) \star T_\l(x)+ T_l(x) \star K
\ee
where $K$ is an $x$-independent operator. By using the boundary condition
for $T_\l$ we obtain
\be
\frac{\partial T_\l(x)}{\partial t}=
V(x)\star T_\l(x)-T_\l(x) \star V(-L) \label{t_l}
\ee
Evaluating \ref{t_l} at $x=L$ and using the boundary condition for 
the operator $V$, we obtain
\be
\frac{\partial T_\l(x)}{\partial t}=[V(L),T_\l(L)]_\star
\ee
Once we have managed to write the time derivative of $T_\l(L)$ as a
commutator, it follows straightforwardly that ${\rm tr}(T_\l(L))$ is
time independent, and we can read-off conserved charges from the
expansion of $T_\l(L)$ in powers of $\l$. In this derivation we had to
use the fact that the $\star$-product is associative and that the
$\star$-inverse of certain operators exist.
 
Before we proceed in constructing the $\star$-zero-curvature condition for 
the sinh-Gordon model we should mention one important aspect: the
noncommutative generalization of a given term is not necessarily unique. 
For example, in going from $\theta=0$ to $\theta \neq 0$ the derivative of
$\phi$ can be written as $\partial \phi$ or as
$\frac{1}{2}(e_\star^{-\phi} \star \partial e_\star^{\phi}-
e_\star^{\phi} \star \partial e_\star^{-\phi})$, and as we will see
later, this ambiguity leads to different equations of motion.

We will work in light-cone coordinates now, where $x_\pm=(x_0 \pm x_1)/2$.  
We can write the equation of motion for the shG model as a zero-condition
equation, by introducing a two component vector potential $A$
and $\bar A$,
\be
A=-\frac{m\lambda}{2}(e^{\beta\phi} \,\, \sigma_- 
+e^{-\beta\phi} \,\, \sigma_+)
\qquad {\rm and} \qquad
{\bar A}=\frac{m}{2\lambda}(\sigma_-+\sigma_+)-
\frac{\beta}{2}\bar\partial \phi \,\, \sigma_3 \label{aabar}
\ee
where $\sigma_{\pm}=\frac{1}{2}(\sigma_1\pm i\sigma_2)$, $\sigma_i$
are the usual Pauli matrices, and $\lambda$ is the spectral parameter.
$A$ and $\bar A$ satisfy
\be
\bar \partial A - \partial \bar A + [A,\bar A]=0 \ , \  \label{zc_lc}
\ee
It is a simple computation to verify that the zero-curvature
condition \ref{zc_lc} with the functions \ref{aabar} are equivalent to
the equation of motion for the shG model \ref{shG}. Notice that in
showing this, the diagonal elements of the matrix equation \ref{zc_lc}
are the equation of motion and the off-diagonal elements vanish.
 
We {\em define} now the noncommutative sinh-Gordon model by the following
$\star$-zero-curvature equation
\be
\bar \partial A- \partial \bar A + [A,\bar A]_{\star}=0 
\ , \ \label{AA}
\ee
where $[A,\bar A]_{\star}=A\star\bar A-\bar A\star A$.  The equation
of motion derived from \ref{AA} is
\be
\partial\bar\partial\phi+\frac{m^2}{\beta}\sinh_\star(\beta\phi)=0 \label{eq1}
\ee
which is exactly the same equation one would obtain from the Moyal deformation
of the shG action. There are, though, two more constraints, coming from the
off-diagonal elements, and which read
\bea
&&\bar \partial (e^{-\beta \phi}_\star)+
\frac{\beta}{2}(e^{-\beta\phi}_\star \star \bar \partial \phi +
                \bar\partial\phi \star e^{-\beta\phi}_\star)=0 \\
&&\bar \partial (e^{\beta \phi}_\star)-
\frac{\beta}{2}(e^{\beta \phi}_\star \star \bar \partial \phi +
\bar\partial \phi \star e^{\beta\phi}_\star)=0 \label{constraints}
\eea
It is easy to show that these constraints can be written as total derivatives,
and that they vanish in the limit $\theta \to 0$.

\subsection{The Grisaru and Penati Proposal}

In \cite{GP} Grisaru and Penati have proposed a system of equations for
the noncommutative sine-Gordon model, using the method of
bicomplexes. We will show that it is possible to obtain their equation
of motion from the $\star$-zero-curvature equation. On the other hand, 
the constraints we find are, apparently, different from theirs.

If we take the gauge potential $A$ and $\bar A$ to be
\bea
&&A=-\frac{m\lambda}{2}(e^{\beta\phi}_{\star}\,\,\sigma_-
                +e^{-\beta\phi}_\star \,\,\sigma_+)
\nonumber \\
&&\bar A=\frac{m}{2\lambda}(\sigma_-+\sigma_+)+\frac{1}{4}
(e^{\beta\phi}_\star \star \bar\partial(e^{-\beta\phi}_\star)-
 e^{-\beta\phi}_\star \star \bar\partial(e^{\beta\phi}_\star)) \,\,\sigma_3
\eea
we obtain the same equation of motion \footnote{After replacing
$\beta$ by $i\beta$.} as in \cite{GP}
\be
\partial(e^{\beta\phi}_\star \star \bar\partial(e^{-\beta\phi}_\star)-
         e^{-\beta\phi}_\star \star \bar\partial(e^{\beta\phi}_\star))=
2 m^2\sinh_\star(\beta \phi)
\ee
but different additional constraints
\bea
&&\bar \partial(e^{-\beta\phi}_\star)-
\frac{1}{4}\{e^{-\beta\phi}_\star, 
e^{\beta\phi}_\star \star \bar\partial(e^{-\beta\phi}_\star)-
e^{-\beta\phi}_\star \star \bar\partial(e^{\beta\phi}_\star) 
\}_\star=0 \\
&&\bar \partial(e^{\beta\phi}_\star)-
\frac{1}{4}\{e^{\beta\phi}_\star, 
e^{-\beta\phi}_\star \star \bar\partial(e^{\beta\phi}_\star)-
e^{\beta\phi}_\star \star \bar\partial(e^{-\beta\phi}_\star)
\}_\star=0 \ . \ 
\eea
It is straightforward to show that these constraints can be written as total
derivatives, and that they vanish in the limit $\theta \to 0$.

\subsection{Euclidean Solitons}

In this section we will study soliton solutions for the ncsG model
defined in the previous sections, equations \ref{eq1} and \ref{constraints}.
We should remark here that by "euclidean solitons" we refer to solutions to
the equations of motion for the ncsG model that correspond to the usual soliton
solution for the sG model, when setting $\theta=0$.

Since the solution for the field $\phi$ itself is a function of the
noncommutative parameter $\theta$, we have to make a double
expansion. Initially we expand the field $\phi$ in a power series in
$\theta$
\be
\phi=\sum_{n=0}^{\infty} \phi_n \theta^n \label{phi}
\ee
We can use this expansion to find the equations of motion and constraints to
first order in $\theta$. Notice that there will be a dependence in $\theta$
arising from the expansion \ref{phi} and also from the definition of
the $\star$-product. We have
\bea
&&\phi^n_{\star}=\phi^n_0+n\phi_0^{n-1}\phi_1\theta+(n\phi^{n-1}_0\phi_2+
\frac{n(n-1)}{2}\phi^{n-2}\phi^{2}_1+ \nonumber \\
&&\phantom{\phi^n_{\star}=}
+\frac{n(n-1)(n-2)}{24}\phi^{n-3}_0 {\cal B}_1 +
\frac{n(n-1)}{8}\phi^{n-2}_0 {\cal B}_2)\theta^2 + o(\theta^3)
\label{expansion1}
\eea
where ${\cal B}_1$ and ${\cal B}_2$ are given by
\bea
&&{\cal B}_1=(\partial\phi_0)^2(\bar{\partial}^2\phi_0)+
  (\bar{\partial}^2\phi_0)(\partial\phi_0)^2-
2\partial\bar{\partial}\phi_0\partial\phi_0\bar{\partial}\phi_0 \nonumber \\
&&{\cal B}_2=\partial^2\phi_0\bar{\partial}^2\phi_0-(\partial \bar{\partial}\phi_0)^2 \label{expansion2}
\eea
See appendix 1 for a derivation of \ref{expansion1} and \ref{expansion2}.
The equations of motion are, to order
$\theta^2$
\bea
&&\partial\bar{\partial} \phi_0 + \frac{m^2}{\beta} \sin(\beta\phi_0)=0 \\
&&\partial\bar{\partial} \phi_1 + m^2 \phi_1 \cos(\beta\phi_0)=0\\
&&\partial\bar{\partial} \phi_2 + m^2 \phi_2 \cos(\beta\phi_0)-
\frac{m^2\beta}{2}\phi_1^2\sin(\beta\phi_0)=0 
\eea
The first two are the same as found by Grisaru and
Penati in \cite{GP}.

The solutions for $\phi_0$, $\phi_1$, and $\phi_2$ 
are readily found, and we refer to
appendix 2 for the details. The solutions are
\bea
&&\phi_0=\frac{4}{\beta} \tan^{-1}(\exp(\frac{m}{\sqrt{1-v^2}} (x-x_0))) 
\\
&&\phi_1=K \phi_0' \\
&&\phi_2=\frac{K^2}{2}\phi_0''
\eea
Where $K$ is a constant of integration. Using these expressions we see
that the series for $\phi(x)$ can be partially resummed to give
$\phi(x+A\theta)$. Actually, it is easy to show that this is indeed the
case to all orders, which establishes the fact that {\em the one soliton
solution for the commutative theory solves the noncommutative equations of
motion}. To show that, we start by establishing that if $f(x_0,x_1)$ and
$g(x_0,x_1)$ depend on their arguments as a linear function of $x_1$ and
$x_2$, say, $x_1-v \, x_0$, then their $\star$-product coincides with
their classical product ($\theta=0$). This can be easily seen by using the
Fourier decomposition of $f$ and $g$,
\bea
&&f \star g=\exp(\frac{i}{2}\theta_{\mu \nu}\partial_x^\mu\partial_y^\nu)
\left.\int dp dq \tilde{f}(p)\tilde{g}(q)\exp(ip(x_1-vx_0)+iq(y_1-vy_0))
\right|_{x=y}= \nonumber \\
&&\phantom{f \star g}=\int dp dq \tilde{f}(p)\tilde{g}(q)
\exp(i(p+q)(x_1-vx_0))=fg \label{product}
\eea
Therefore the $\sin(\beta\phi)_\star=\sin(\beta\phi)$ and the equation
of motion turns out to be the same as in the usual sine-Gordon model.
The next step to be verified is to check if the constraints are
satisfied.  This again is easily shown to be the case, since from
\ref{product} we can take the constraints to be evaluated at
$\theta=0$, when they become trivial.

The study of multi-soliton solutions is not as simple as the one-soliton
case, and we shall not pursue it here.
 
\subsection{Noncommutative Principal Chiral Model}

In the previous subsections we studied models where the na\"\i ve
noncommutative version fails to be integrable. In this subsection we
will study a model where the na\"\i ve construction works. This is the
$U(N)$ principal chiral model (pcm).

The action of the $U(N)$ pcm is 
\be
S_{pcm}=\frac{1}{2g^2_0}\int d^2x {\rm Tr}
(\partial_{\mu}g^{-1}\partial^{\mu} g) \label{s}
\ee 
where $g$ takes values in $U(N)$. The equation of motion
of the pcm is easily seen to be $\partial_\mu (g^{-1}\partial^{\mu}g)=0$.
The Moyal deformation of the pcm is given by the action
\be
S^*_{pcm}=\frac{1}{2g^2_0}\int d^2x {\rm Tr}
(\partial_{\mu}g^{-1} \star \partial^{\mu} g) \label{s*}
\ee
and the field $g$ is required to satisfy $g \star g^{\dagger}= g^{\dagger}\star
g =1$. 
The reason why we should be specific about the group to which $g$
belongs is that not all groups allow noncommutative extensions, for
example, there is no noncommutative $SU(N)$. Therefore we will
restrict our analysis to the $U(N)$ pcm. This model was recently
studied in \cite{P}. The restriction to $U(N)$ has also been shown
to be of great importance for renormalization requirements \cite{A}.

Notice that, as we mentioned earlier, the quadratic part of a
noncommutative action in a manifold without boundaries is equivalent
to the commutative action. Therefore the actions \ref{s} and \ref{s*}
are equivalent \footnote{Provided the field $g$ falls off fast enough
at infinity.}. The main difference between the commutative and
noncommutative models relies therefore not in their actions, but in
the constraints that the field $g$ satisfies.

We would like to construct nontrivial conserved charges for this
model, in order to show that it is integrable. We can do so by
following the Brezin-Itzykson-Zinn-Justin-Zuber (BIZZ) method
\cite{BIZZ}, which is extremely simple and yet powerful. For the sake
of completeness we summarize the BIZZ construction here.

In \cite{BIZZ} BIZZ start by assuming that there exists a set of
matrices $A_{\mu}^{\alpha \beta}$ satisfying the following conditions:
\begin{itemize}
\item
The field $A_\mu$ is a pure gauge, that is, one can find a nonsingular
matrix $h$ such that $A_\mu=h^{-1}\partial_\mu h$
\item
As a consequence of the equations of motion we should have
$\partial_\mu A_\mu=0$
\end{itemize}
Based on these two requirements, it can be shown that the recursively
defined currents $J_\mu^{(n+1)}=D_\mu \chi^{(n)}$, $n \geq 0$, are
conserved, where $D_\mu^{\alpha \beta}=\delta^{\alpha
\beta}\partial_\mu + A_{\mu}^{\alpha \beta}$ is a covariant derivative
satisfying the zero-curvature condition 
$[D_\mu,D_\nu]=0$, and the fields $\chi^{(n)}$ are defined
by $J_\mu^{(n)}=\epsilon_{\mu \nu} \partial_\nu \chi^{(n)}$, $n \geq
1$, and we start with $\chi^{(0)}=1$. It is a simple exercise to show
that $\partial_\mu J_\mu = 0$. By construction, these are nonlocal charges.

This construction was used by BIZZ to establish the integrability of
the pcm, since we can take $A_\mu=g^{-1}\partial\mu g$, as we see from
the equation of motion of the pcm, which automatically satisfies both
requirements stated above. In order to carry out the
derivation, though, we have to establish one crucial point: the Moyal
deformed commutator $[D_\mu,D_\nu]_\star = D_\mu \star D_\nu - D_\nu
\star D_\mu$, should vanish.  This is easily seem to be the case,
bearing in mind that the identities $\partial_\mu g^{-1} \star g=
-g^{-1} \star \partial_\mu g$ and so on, are still valid. One such
conserved charge that we can write is 
\be 
{\cal Q}^{(2)}=-\int_{-\infty}^{+\infty}dx\,\, g^{-1}\star \partial_1 g
+\int_{-\infty}^{+\infty}dx \,\, g^{-1}\star \partial_0 g \star
\int_{-\infty}^{x} d{x'} \,\, g^{-1}\star \partial_0 g 
\ee

\subsection{Zero-Curvature Equation for the Noncommutative PCM}

We can also write the noncommutative equation of motion of the
$U(N)_\star$ pcm as a zero-curvature condition. Consider the
potentials
\bea
&&U(\lambda)=\frac{1}{2}\frac{l_0+l_1}{1-\lambda}-
\frac{1}{2}\frac{l_0-l_1}{1+\lambda} \\
&&V(\lambda)=\frac{1}{2}\frac{l_0+l_1}{1-\lambda}+
\frac{1}{2}\frac{l_0-l_1}{1+\lambda}
\eea
where
\be
l_0(x,t)=\frac{\partial g}{\partial t}*g^{-1}
\qquad {\rm and} \qquad
l_1(x,t)=\frac{\partial g}{\partial x}*g^{-1}
\ee
Introducing this on the zero-curvature condition \ref{zc} we obtain
\be
\frac{\partial^2 g}{\partial t^2}-\frac{\partial^2 g}{\partial x^2}=
\frac{\partial g}{\partial t}*g^{-1}*\frac{\partial g}{\partial t}-
\frac{\partial g}{\partial x}*g^{-1}*\frac{\partial g}{\partial x}
\ee
which is the equation of motion of the pcm, and it can be rewritten in
the more compact form
\be
\partial_{\mu}(g^{-1}*\partial^{\mu}g)=0 \ . \
\ee
Contrary to the noncommutative sG model, there are no further constraints
in the noncommutative $U(N)$ pcm.
 
\section{Conclusions}
We have seen that the Moyal deformation of a given 2d integrable model does 
not necessarily provide a integrable field theory. In the case of the 
sinh-Gordon model (and by replacing $\beta \to i\beta$, the sine-Gordon model) 
we were able to establish their {\em non}-integrability by
computing the amplitude for $2 \to 4$ particles at the tree-level, and verifying it is non-zero. 
On the other hand, the noncommutative $U(N)$ principal chiral model defined through the Moyal
deformation of the action and constraints of the $U(N)$ principal chiral model, does provide an
integrable field theory, where the elegant method of Brezin et al \cite{BIZZ}
works as well as in the commutative case.

The equations of motion we found initially \ref{eq1} and \ref{constraints} are different from the
ones proposed by Grisaru and Penati in \cite{GP}. Upon a change in the definition of the noncommutative
version of $\partial \phi$ we were able to find the same equation of motion as \cite{GP}, 
but it is not trivial to establish the equality of the constraints.

By looking at the equations of motion in a perturbative form, we found 
the "euclidean solitons" for the noncommutative sine-Gordon model, and showed that 
the 1-soliton solution of the sine-Gordon model solves the equations of motion and constraints 
of the noncommutative version. 

There are several interesting directions to pursue. Initially, it would be nice to have a more thorough
understanding of the conservation laws, verifying for example, that these charges are in involution. Next
one could consider the noncommutative versions of different models, such as the affine Toda theories. And last but not least, the investigation of the quantization of these models is a fascinating, if somewhat difficult problem.

\section*{Acknowledgments}

We would like to thank the hospitality of the Abdus Salam ICTP, where
we started this work. Discussions with P. Dorey, J. Evans, A. Garcia,
K. Narayan, A. Leclair, M. Neubert, V. Sahakian, M.M. Sheikh-Jabbari, and
J. Varilly are gracefully acknowledged. One of us (MM) would like to thank 
F. Muller-Hoisen for email exchanges, and  C. Wotzasek for the use
of computer facilities. This work is in part supported
by the NSF and CNPq (profix) (MM) and Fapesp (ICC).

\newpage
\appendix

\section{Expansion of $\phi^n_\star$ to order $\theta^2$}

We want to find $\phi^n_\star$ to order $\theta^2$, where 
$\phi=\sum_{n=0}^{\infty}\phi_n\theta^n$. We write $\phi^n_\star$
as
\be
\phi^n_\star=\phi_0^n+\theta A_n + \theta^2 B_n + o(\theta^3) \ . \
\ee
Using the associativity of the $\star$-product, 
$\phi^{n+1}_\star=\phi^n_\star \star \phi$, we find the following recurrence
equation for $A_n$
\be
A_{n+1}=A_n\phi_0+\phi^n_0\phi_1
\ee
which is readily solved by $A_n=n\phi^{n-1}_0\phi_1$. For $B_n$ it is 
convenient to introduce
\be
B_n=\alpha_n\phi^{n-2}_0\phi^2_1+\beta_n\phi^{n-1}_0\phi_2+
\gamma_n\phi^{n-3}_0 {\cal B}_1+\delta_n \phi^{n-2}_0 {\cal B}_2
\ee
where ${\cal B}_1=(\partial\phi_0)^2
{\bar\partial}^2\phi_0+({\bar\partial}\phi_0)^2 \partial^2\phi_0-2\partial\phi_0{\bar\partial}\phi_0\partial\bar\partial\phi_0$ and
${\cal B}_2=\partial^2\phi_0{\bar\partial}^2\phi_0-
(\partial\phi_0{\bar\partial}\phi_0)^2$. We find the following
recurrence relations
\bea
&&\alpha_{n+1}=\alpha_n+n \nonumber \\
&&\beta_{n+1}=\beta_n+1 \nonumber \\
&&\gamma_{n+1}=\gamma_n+\frac{n(n-1)}{8} \nonumber \\
&&\delta_{n+1}=\delta_n+\frac{n}{4}
\eea
which are solved by $\alpha_n=n(n-1)/2$, $\beta_n=n$, 
$\gamma_n=n(n-1)(n-2)/24$, and $\delta_n=n(n-1)/8$.
The final form, is then
\bea
\phi^n_\star=\phi^n_0+n\phi^{n-1}_0\phi_1\theta+
\theta^2 (n\phi^{n-1}_0\phi_2+\frac{n(n-1)}{2}\phi^{n-2}_0 \phi_1^2 ) \\ \nonumber
+\frac{n(n-1)}{8}\phi^{n-2}_0 {\cal B}_{\bf 2 } \theta^2
+\frac{n(n-1)(n-2)}{24}\phi_0^{n-3}{\cal B}_1{\bf \theta^2}
\eea
Notice that ${\cal B}_1$ and ${\cal B}_2$ vanish for soliton solutions.

\section{Solution of the Equation of Motion to Order $\theta^2$}

In section 3.4 we found the classical equations of motion for the
noncommutative sine-Gordon model. Taking $\beta=1$, their static form to order
$\theta^0$, $\theta^1$, and $\theta^2$, are
\bea
&&\phi_0^{''}=C\sin\phi_0 \label{eq_0}\\
&&\phi_1^{''}=C\phi_1\cos\phi_0 \label{eq_1}\\
&&\phi_2^{''}=C\phi_2 \cos \phi_0 -\frac{C}{2}\phi_1^2 \sin \phi_0 
\label{eq_2}
\eea
where the primes indicate space derivatives, and $C$ is a constant 
($C=\frac{m^2}{1-v^2}$, and $v$ is the soliton velocity). To solve the 
first equation \ref{eq_0}
we multiply it by $\phi_0^{'}$ and integrate it, using the boundary conditions
$\phi_0=0$ at $x=0$ and $\phi_0=2\pi$ for $x \to \infty$
\be
\phi_0^{'}=\pm 2 \sqrt{C}\sin\frac{\phi_0}{2} \label{eq_01}
\ee
The plus (minus) sign corresponds to the soliton (anti-soliton) solution.
The solution of \ref{eq_01} is easily found to be
\be
\phi_0=4\tan^{-1}(\exp(\sqrt{C}(x-x_0)))
\ee
In order to solve \ref{eq_1} we multiply it by $\phi_0^{'}$ and use
\ref{eq_0} to obtain
\be
\phi_0^{'}\phi_1^{''}=C\phi_1\phi_0^{'}\cos\phi_0=\phi_1\frac{d}{dx}
(C\sin\phi_0)=\phi_1\phi_0^{''}
\ee
and the last equation together with the fact that the derivatives of
$\phi$ vanish for $x \to \infty$ implies that
\be
\phi_1\phi_0^{''}-\phi_1{'}\phi_0^{'}=0
\ee
This equation is easily solved by
\be
\phi_1=\frac{K}{\cosh(\sqrt{C}(x-x_0))}
\ee
where $K$ is an integration constant. The solution for $\phi_2$ can be found in the following way:
take two derivatives of \ref{eq_0} and multiply it by $\frac{K^2}{2}$, to get
\be
(\frac{K^2}{2}\phi_0^{''})^{''}=C(\frac{K^2}{2}\phi_0^{''})\cos\phi_0-
\frac{C}{2}(K\phi_0^{'})^2 \sin\phi_0
\ee
which is the same as \ref{eq_2}. Since this is a first order equation for $\phi_2$ and the boundary
conditions at infinity are satisfied automatically, we can write the solution for $\phi_2$ as
\be
\phi_2=\frac{K^2}{2}\phi_0^{''}
\ee
This solves the equations of motion for the ncsG model o second order in $\theta$.


\newpage
\begin{thebibliography}{99}

\bibitem{SW}
N.~Seiberg and E.~Witten,
``String theory and noncommutative geometry'',
JHEP {\bf 9909}, 032 (1999) [arXiv:hep-th/9908142].
%%CITATION = HEP-TH 9908142;%%

\bibitem{MvRS}
S.~Minwalla, M.~Van Raamsdonk and N.~Seiberg,
``Noncommutative perturbative dynamics'',
JHEP {\bf 0002}, 020 (2000)
[arXiv:hep-th/9912072].
%%CITATION = HEP-TH 9912072;%%

\bibitem{Suss}
L.~Susskind,
``The quantum Hall fluid and non-commutative Chern Simons theory'',
arXiv:hep-th/0101029.
%%CITATION = HEP-TH 0101029;%%

\bibitem{GM}
J.~Gomis and T.~Mehen,
``Space-time noncommutative field theories and unitarity'',
Nucl.\ Phys.\ B {\bf 591}, 265 (2000)
[arXiv:hep-th/0005129].
%%CITATION = HEP-TH 0005129;%%

\bibitem{CLZ}
C.~S.~Chu, J.~Lukierski and W.~J.~Zakrzewski,
``Hermitian analyticity, IR/UV mixing and unitarity of noncommutative
field theories'',
Nucl.\ Phys.\ B {\bf 632}, 219 (2002)
[arXiv:hep-th/0201144].
%%CITATION = HEP-TH 0201144;%%

\bibitem{BDFP}
D.~Bahns, S.~Doplicher, K.~Fredenhagen and G.~Piacitelli,
``On the unitarity problem in space/time noncommutative theories'',
Phys.\ Lett.\ B {\bf 533}, 178 (2002)
[arXiv:hep-th/0201222].
%%CITATION = HEP-TH 0201222;%%

\bibitem{FT}
L.~D.~Faddeev and L.~A.~Takhtajan,
``Hamiltonian methods in the theory of solitons'',
Berlin, Germany: Springer (1987) 592 p. 
(Springer Series In Soviet Mathematics).

\bibitem{BIZZ}
E.~Brezin, C.~Itzykson, J.~Zinn-Justin, and J.-B.~Zuber,
``Remarks about the existence of non-local charges in 
two-dimensional models'',
Phys.\ Lett.\ B{\bf 82},442-444 (1979).
%%CITATION = PHLTA,B82,442;%%

\bibitem{reviews}
M.~R.~Douglas and N.~A.~Nekrasov,
``Noncommutative field theory'',
Rev.\ Mod.\ Phys.\ {\bf 73}, 977 (2001)
[arXiv:hep-th/0106048].
%%CITATION = HEP-sH 0106048;%%
\\
R.~J.~Szabo,
``Quantum field theory on noncommutative Spaces'',
arXiv:hep-th/0109162.
%%CITATION = HEP-TH 0109162;%%

\bibitem{Filk}
T.~Filk,
``Divergencies in a field theory on quantum space'',
Phys.\ Lett.\ B {\bf 376}, 53-58 (1996).
%%CITATION = PHLTA,B376,53;%%

\bibitem{ZZ}
A.~B.~Zamolodchikov and A.~B.~Zamolodchikov,
``Factorized S-matrices in two dimensions as the exact solutions of certain relativistic quantum field models'',
Annals Phys.\  {\bf 120}, 253 (1979).
%%CITATION = APNYA,120,253;%%

\bibitem{D}
P.~Dorey,
``Exact S-matrices'',
arXiv:hep-th/9810026.
%%CITATION = HEP-TH 9810026;%%

\bibitem{NOS}
C.~Nunez, K.~Olsen, and R.~Schiappa,
``From noncommutative bosonization to S-duality'',
JHEP {\bf 0007}, 030 (2000).
%%CITATION = HEP-TH 0005059;%%

\bibitem{P}
S.~Profumo,
``Noncommutative principal chiral models'',
arXiv:hep-th/0111285.
%%CITATION = HEP-TH 0111285;%%

\bibitem{LP}
O.~Lechtenfeld, A.~D.~Popov,
``Noncommutative multi-solitons in 2+1 dimensions'',
JHEP {\bf 0111}, 040 (2001)
[arXiv:hep-th/0106213].
%%CITATION = HEP-TH 0106213;%%

\bibitem{Sh}
R.~Shankar,
``A Model that acquires integrability and O(2n) 
invariance at a critical coupling'',
Phys.\ Lett.\ B {\bf 102}, 257 (1981).
%%CITATION = PHLTA,B102,257;%%

\bibitem{GP}
M.~T.~Grisaru and S.~Penati,
``The noncommutative sine-Gordon system'',
arXiv:hep-th/0112246.
%%CITATION = HEP-TH 0112246;%%

\bibitem{A}
A.~Armoni,
``Comments on perturbative dynamics of noncommutative Yang-Mills 
theory'''
Nucl.\ Phys.\ B {\bf 593}, 229 (2001)
[arXiv:hep-th/0005208].
%%CITATION = HEP-Th 0005208;%%

\end{thebibliography}
\end{document}